# Chern Insulators, van Hove singularities and Topological Flat-bands in Magic-angle Twisted Bilayer Graphene

Shuang Wu[1†], Zhenyuan Zhang[1†], K. Watanabe[2], T. Taniguchi[2], and Eva Y. Andrei[1*]

[1]Department of Physics and Astronomy, Rutgers University, New Jersey, 08854, USA

[2]National Institute for Materials Science, Namiki 1-1, Tsukuba, Ibaraki 305 0044, Japan

† These authors contributed equally to the work.

* Corresponding author. Email: andrei@physics.rutgers.edu

**Magic-angle twisted bilayer graphene (MA-TBG) exhibits intriguing quantum phase transitions triggered by enhanced electron-electron interactions when its flat-bands are partially filled. However, the phases themselves and their connection to the putative non-trivial topology of the flat bands are largely unexplored. Here we report transport measurements revealing a succession of doping-induced Lifshitz transitions that are accompanied by van Hove singularities (VHS) which facilitate the emergence of correlation-induced gaps and topologically non-trivial sub-bands. In the presence of a magnetic field, well quantized Hall plateaus at filling of 1, 2, 3 carriers per moiré-cell reveal the sub-band topology and signal the emergence of Chern insulators with Chern-numbers, $C = 3, 2, 1$, respectively. Surprisingly, for magnetic fields exceeding 5T we observe a VHS at a filling of 3.5, suggesting the possibility of a fractional Chern insulator. This VHS is accompanied by a crossover from low-temperature metallic, to high-temperature insulating behavior, characteristic of entropically driven Pomeranchuk-like transitions.**

The band structure of twisted bilayer graphene is strongly renormalized by the moiré superstructure.[1-9] Close to the 'magic' twist-angle, $\theta \sim 1.1°$, the low-energy electronic structure of MA-TBG consists of two four-fold degenerate nearly-flat bands straddling the charge neutrality point (CNP) and separated from higher-energy bands by spectral gaps (ED-Fig. 1). Aligning the Fermi-level with these flat bands, facilitates interaction-induced instabilities at integer moiré-cell fillings leading to strong correlations[7,8,10-14] and to the creation of topologically non-trivial flavor-polarized (spin/valley) moiré sub-bands[15,16] characterized by finite Chern-numbers. In the absence of broken $C_{2z}$ $\mathcal{T}$ symmetry ($C_{2z}$ = 180° in plane rotation and $\mathcal{T}$ = time reversal), the bands are degenerate and their non-trivial topology is hidden. Lifting this degeneracy, by breaking time-reversal or sublattice symmetry,[17,18] is expected to reveal the band-topology and its role in shaping the correlated phases, but thus far experimental evidence is limited. Here, by applying a magnetic-field to break the time-reversal symmetry, we uncover sub-bands with non-trivial topology at integer fillings, whose emergence is triggered by a succession of Stoner-like instabilities and facilitated by van Hove singularities (VHS). The sub-band topology is revealed by the finite Chern numbers obtained from the observation of quantized Hall plateaus.

**Full and half-full Landau fans at integer moiré fillings**

Transport measurements were carried out on a MA-TBG sample with twist angle $\theta \approx 1.17° \pm 0.02$ (Fig. 1a) (Methods). At low temperatures the longitudinal resistance, $R_{xx}(n)$, develops strong peaks at integer moiré-cell fillings, $|n/n_0| = 0, 2, 3$, and diverges at the band edges, $|n/n_0| = 4$, (Fig. 1b).[7] Here $n$ is the gate-controlled carrier density, $n/n_0$ is the number of carriers per moiré-cell (moiré filling factor), $n_0 = 2/\sqrt{3}/(\theta/a)^2$ corresponds to one carrier per moiré-cell, $a$ =0.246nm is graphene's lattice constant, and $\theta$ is the twist-angle in radians. Close to $n/n_0 = -2$ (2 holes per moiré-cell) the sample is superconducting with a critical temperature of $T_c \sim 3.5K$ (ED-Fig. 2), consistent with earlier reports.[8,10] Since superconductivity is not the focus of this work all measurements were carried out at magnetic fields or driving currents where superconductivity is quenched.

The density and magnetic-field ($B$) dependence of $R_{xx}$ (Fig. 1c) features Shubnikov de Haas minima (SI) resulting in Landau fans that are controlled by the competing periods of the moiré lattice and the magnetic-flux array. This produces fans whose pleats (trajectories) can be parameterized according to the partial Diophantine equation: $n/n_0(s,v) = s + v\,(\phi/\phi_0)$ where $s, v \in \mathbb{Z}$ and $s = 0$ or $s \cdot v > 0$. Here $\phi$ is the magnetic-flux per moiré-cell, $\phi_0 = h/e$ is the magnetic-flux quantum, $h$ is Planck's constant, $v$ is the flux filling factor representing the number of carriers per flux line, and $s$ is the moiré filling factor (branch index) corresponding to the number of carriers per moiré-cell in zero field. Trajectories emanating from the CNP, $s = 0$, which are unaffected by the moiré potential form a bilateral (full) Landau fan, ($v = \pm 1, \pm 2, \pm 3, \pm 4, \pm 8, \pm 12 \ldots$ for B > 2T) characteristic of LLs in a translationally invariant system where both electron and hole carriers contribute to the fan. By contrast, trajectories on the $s \neq 0$ branches form unilateral (half) Landau fans that slope away from the CNP. These trajectories resemble the non-interacting Hofstadter butterfly spectrum,[19] but the underlying physics is different. The half-fans observed here, are a direct consequence of correlation-induced spectral gaps emerging on the $s \neq 0$ branches after the doping level has reached the corresponding integer moiré filling, while they are absent on the low filling side of the branch. This is distinctly different from the bilateral Hofstadter butterflies arising from rigid Bloch bands where the gaps are controlled by the lattice period and independent of filling.

From $R_{xx}$ and $R_{xy}$, we obtain the conductivities, $\sigma_{xy} = R_{xy}/(R_{xy}^2 + (w/l)^2 R_{xx}^2)$ and $\sigma_{xx} = (w/l)R_{xx}/(R_{xy}^2 + (w/l)^2 R_{xx}^2)$, where $w/l$ =0.625 is the sample aspect ratio (Fig. 1d and ED-Fig. 3). To avoid artifacts associated with lead asymmetry, the resistances are symmetrized: $\bar{R}_{xx}(B) = (R_{xx}(n, B) + R_{xx}(n, -B))/2$ and $\bar{R}_{xy}(B) = (R_{xy}(B) - R_{xy}(-B))/2$, and henceforth labeled $R_{xx}$ and $R_{xy}$ respectively. The high quality of this sample is reflected in the appearance of well-quantized $\sigma_{xy}$ plateaus and $\sigma_{xx}$ minima on the $s = 0$ branch. At $B \sim 1.5$T, the $s = 0$ bilateral fan develops a quantized Hall plateau sequence $\sigma_{xy} = v\,e^2/h$, initially with $v = \pm 4, \pm 8$ reflecting the 4-fold degeneracy of the bands (SI). At higher fields, $B > 5$T, the addition of all odd-index plateaus to the sequence $v = 0, 1, \pm 2, \pm 3, \pm 4$ indicates fully lifted spin and valley degeneracies (Fig. 1d).

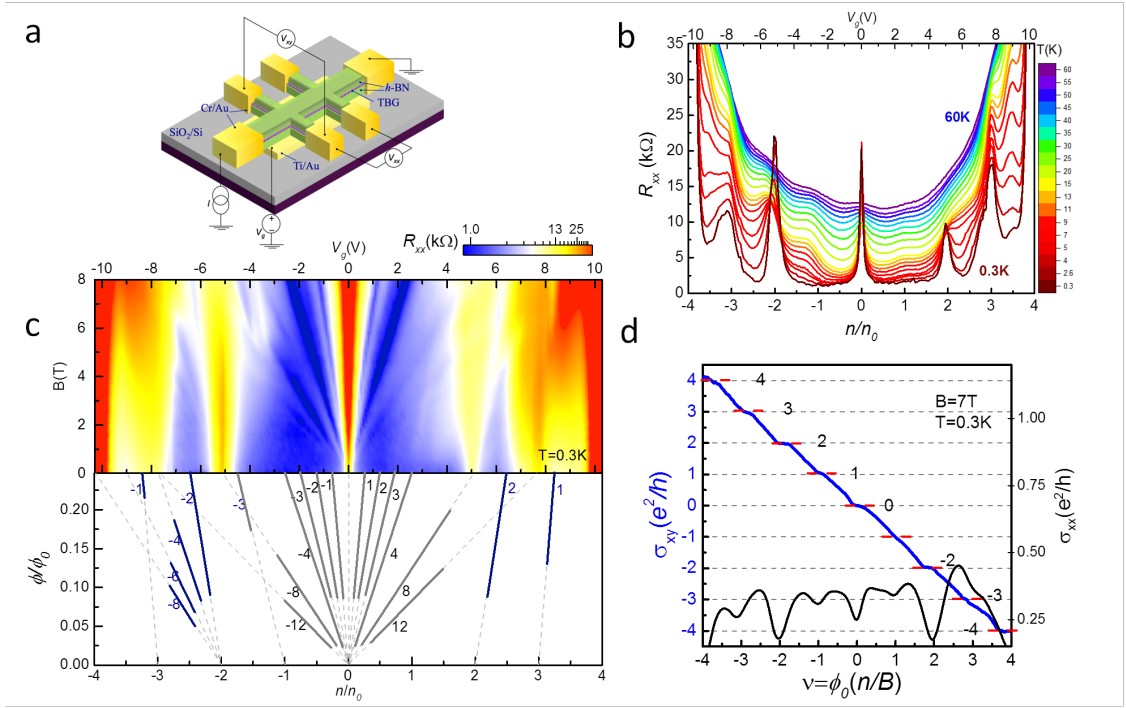

**FIG. 1. Correlated states, Landau fans in MA-TBG (a)** Schematic diagram of multi-terminal Hall bar device for transport measurements on MA-TBG (purple) encapsulated in hBN (light-blue) with Cr/Au edge contacts and a local Ti/Au back-gate for applying the gate voltage ($V_g$). The longitudinal ($V_{xx}$) and transverse ($V_{xy}$) voltages are measured in response to the applied AC current (I). **(b)** Temperature dependence of the longitudinal resistance, $R_{xx}$, versus moiré filling ($n/n_0$) at $B = 0$T from 60K down to 0.3K. The driving current is 100nA. **(c)** Top panel: diagram of $R_{xx}(B, V_g)$. Bottom panel: Landau fans are parameterized by their pleat and branch index $v, s \in \mathbb{Z}$ according to the Diophantine equation $v(\phi/\phi_0) = n/n_0(s, v) - s$, where $\phi/\phi_0$ is the number of flux lines per moiré cell. Gray lines trace the $s = 0$ full-Landau fan, and blue lines trace the $s \neq 0$ half-Landau fans. **(d)** $\sigma_{xy}$ and $\sigma_{xx}$ as a function of Landau level filling in the $s = 0$ branch at $B = 7$T, $T = 0.3$K. The integer quantum Hall plateaus $\sigma_{xy} = v\, e^2/h$ (dashed red lines) and minima of $\sigma_{xx}$ at $v = 0, 1, \pm 2, \pm 3, \pm 4$, indicate twist-angle homogeneity across the sample and its boundaries, which is key to probing the Fermi-surface topology.

**Van Hove singularities and band reconstruction probed by the Hall density**

Electron-electron interactions leading to complex quantum phases, are significantly enhanced by density of states (DOS) peaks, such as VHS, where the Fermi-surface topology changes. Experimentally, such changes known as Lifshitz transitions,[20,21] can be inferred from the Hall density, $n_H = -B/(eR_{xy})$, ($e$ is the electron charge). For a clean 2D system with closed Fermi pockets, $n_H = DA_{FS} = n$, where $D$ is the degeneracy and $A_{FS}$ is the net area enclosed by the Fermi-surface. Thus, far from Lifshitz transitions, $n_H$ measures the free carrier density which determines transport properties. Upon approaching a VHS, $n_H$ diverges logarithmically with opposite signs on the low and high density sides of the VHS.[22] However if the bands become malleable, as often happens when the Fermi-level approaches a VHS, this is no longer the case. For example, if a gap opens upon crossing the VHS, then $n_H$ resets to zero in the newly created empty band. Beyond this point, $n_H$ still increases linearly with $n$, but with an offset: $n_H = n - n_c$, where $n_c$ marks the density where the gap opened. Spectral gaps emerging in the absence of VHS, for example by magnetically enhanced interactions, can generate a similar offset in $n_H$, but without being preceded by a logarithmic divergence. Thus, the evolution of $n_H$ with doping provides access to the Fermi-surface reconstruction and to the emergence of broken symmetry states as the Fermi-level is swept across the band[22].

The doping dependence of $n_H$ was obtained from measurements of $R_{xy}(n)$ (Fig. 2a,b). For $0 \leq |n/n_0| < 2$ we find $n_H \approx n$ (Fig. 2b), indicating closed Fermi-pockets. Upon approaching $|n/n_0| = 2$ from the low-density side, the strong deviation from $n_H \approx n$ fits the expression describing the Lifshitz transition at a VHS in the low-field limit (ED-Fig. 4,5)[22], $n_H \approx \alpha + \beta(n - n_c)\ln|(n - n_c)/n_0|$, $\alpha$, $\beta$ are constants, and $n_c = \pm 2n_0$. The non-interacting, rigid, band structure of MA-TBG contains two VHS separating a two-pocket Fermi-surface centered on the corners of the moiré Brillouin-zone at $|n/n_0| < 2$ from a higher energy single pocket Fermi-surface centered on the $\Gamma$ point at $|n/n_0| > 2$ (insets Fig. 2b).[23] However, the logarithmic divergence of $n_H$ expected to appear on the high-density side of $n_c$ for a VHS in a rigid band structure, is missing. Instead, upon crossing the VHS at $n/n_0 = +2$ ($n/n_0 = -2$), $n_H$ resets to zero and subsequently increases linearly, $n_H = n - 2n_0$ ($n_H = n +$

$2n_0$), but with an offset of $2n_0$ corresponding to a degeneracy reduction from 4 to 2. This transition reflects a correlation-induced gap which triggers a band-occupancy redistribution analogous to the Stoner mechanism. As a result, the initial evenly populated flavor-degenerate four bands transition to two fully-occupied low-energy sub-bands separated by the gap from two higher energy empty sub-bands. Doping the sample beyond this level starts filling the higher energy sub-bands which again consist of closed Fermi pockets. This band reconstruction explains the appearance of the half-Landau fan on the high-density sides of $|s|=2$ and its absence on the low-density sides.

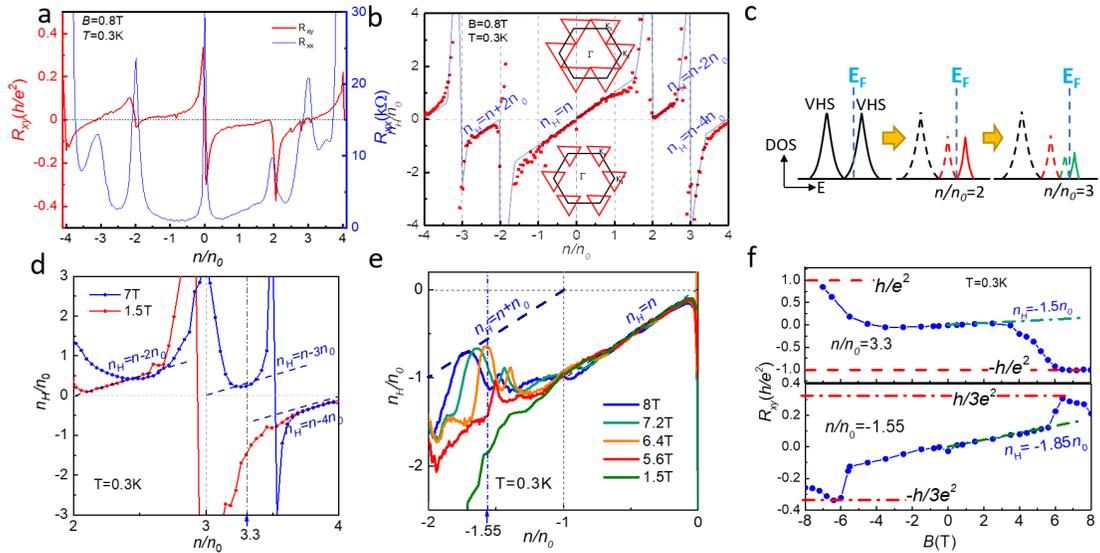

**FIG. 2. Van Hove singularities and correlation-induced sub-bands (a)** Doping dependence of longitudinal ($R_{xx}$) and Hall resistance ($R_{xy}$) at $B = 0.8$T, $T = 0.3$K. **(b)** Doping dependence of Hall density $n_H = -(1/e)(dR_{xy}/dB)^{-1}$ (symbols) together with fits (solid lines) to the logarithmic divergence characteristic of VHS in the vicinity of $|n/n_0| = 2, 3$. Insets: schematic evolution of the Fermi-surface (red triangles) from separate pockets centered on the $K_s$ points in the mini-Brillouin zone for low doping (bottom) and merging together at the VHS (top) close to $|n/n_0| = 2$. **(c)** Diagram depicting scenario for the evolution of the DOS with Fermi-level. Crossing a VHS facilitates a Stoner instability which opens a gap creating a new sub-band with a VHS at its center. **(d)** Doping dependence of reduced Hall density, $n_H/n_0$, near $n/n_0 = 3$. The arrow marks the density for the data in the top panel of (f). Slopes corresponding

to the density dependence of $n_H$ in each branch are marked by dashed lines. The appearance of a VHS at $n/n_0 = 3.5$ in high field gives rise to the divergent Hall density on both sides of this filling. **(e)** Evolution of the $n_H/n_0$ (n) curves with magnetic field near $n/n_0 = -1$. The slope change from $n_H = n$ at low fields ($B = 1.5$T) to $n_H = n + n_0$ for $B > 5.6$T reveals a new sub-band created by a magnetically induced correlation gap on the $s = -1$ branch. **(f)** Top: $R_{xy}(B)$ at $n/n_0 = 3.3$. The low field slope of $R_{xy}(B)$ (blue dashed line) corresponding to $n_H$, abruptly changes sign at ~3.5T indicating the formation of a new sub-band. Bottom: $R_{xy}(B)$ at $n/n_0 = -1.55$ and $T = 0.3$K. The low-field slope of $R_{xy}(B)$ (blue dashed line) abruptly changes at ~6T indicating the emergence of a magnetically induced gap on the s = $-1$ branch.

As doping approaches $|n/n_0| = 3$, the Hall density deviates from $n_H = n \pm 2n_0$ and again displays the logarithmic divergence characteristic of a VHS (solid line in Fig. 2b). The fact that at low fields this divergence appears on both sides, of $|n_c/n_0|$~3, together with the absence of a Landau fan, indicates that no gap opens when crossing this VHS. The approach to this VHS is accompanied by a strong $R_{xx}$ peak at $|n/n_0|$~3 (Fig. 2a), consistent with the singular DOS and concomitant suppression of the Fermi velocity expected at a VHS.[24,25] The logarithmic divergence of $n_H$ and its appearance in tandem with the peak in $R_{xx}$, provide a unique fingerprint enabling us to identify the VHS in this system.

With increasing magnetic field this picture changes qualitatively. As shown in Fig. 2d, the Hall-density around the VHS at $n/n_0 = 3$ at 1.5T is very different from that 7T. At 7T the logarithmic divergence on the high doping side of the VHS resets to zero and then grows linearly, $n_H = n - 3n_0$ but with an offset of $3n_0$, indicating the creation of a new sub-band with lifted degeneracy. As before, when the initially degenerate two sub-bands are half filled, a correlation induced Stoner-like instability that is facilitated by the VHS at $n/n_0 = 3$, splits them into a fully occupied lower energy sub-band that is separated by a gap from a higher-energy empty sub-band. A more detailed perspective on this transition is gained from the field dependence of $R_{xy}$(B) (Fig. 2f top) measured at a fixed density $n/n_0 = 3.3$ (blue arrow in Fig. 2d). For $|B| < 3$T, $R_{xy}$ grows linearly with field, with a slope corresponding to $n_H = -1.5n_0$, indicating hole carriers.

The deviation from the value expected for a closed Fermi-surface at this filling, $n_H = n - 4n_0 = 0.7n_0$, reflects the proximity to the VHS at $n/n_0 = 3$ as is clearly seen in Fig. 2d. Beyond $|B|\sim 3$T, a change in slope, together with a sign change in $n_H$ signals the formation of a correlation-induced gap on the $s = +3$ branch and the appearance of electron-like charge-carriers near the bottom of the newly created sub-band.

Surprisingly, concomitant with the gap opening on the s=3 branch, yet another VHS is born at $n/n_0 \sim 3.5$ as indicated by the logarithmically divergent $n_H$ on both sides of this filling (Fig. 2d and ED-Fig. 6). The pronounced $R_{xx}$ peak at this filling (Fig 4c inset and ED-Fig. 10) is consistent with the singular DOS and simultaneous suppression of the Fermi velocity at a VHS. The band reconstruction and succession of transitions in response to doping[26,27] which is schematically summarized in Fig. 2c, highlights the role of the VHS in facilitating the creation of correlation-induced gaps and the genesis of sub-bands that in turn harbor VHS at their center.

For $s = |1|$, no VHS is available to enhance *e-e* interactions and facilitate the emergence of a gap. As a result, we observe no gap on the $s = +1$ branch, while on the $s = -1$ branch a gap appears only above 6T (Fig. 2e, 2f bottom) as signaled by the slope change of the Hall-density from $n_H = n$ to $n_H = n + n_0$.

As doping approaches the band-edges, $|n/n_0| = 4$, (Figs. 2b, 2d), we observe $|n_H| = 4n_0 - |n|$ at all fields consistent with the theoretical prediction of a single hole (electron) pocket Fermi-surface centered on the Γ-point in the completely full (empty) band.[23]

**Pseudo-Landau levels, quantum Hall plateaus and Chern insulators**

The flat bands in MA-TBG can be mapped onto eight degenerate zeroth order pseudo Landau levels (pLL) corresponding to Dirac fermions coupled with opposite pseudo magnetic fields, opposite chiralities and opposite valleys.[15] This model reflects a gauge-field generated by the moiré potential which produces pseudo-magnetic fields with opposite signs in the K and K' valleys: $B^+_{PMF}, B^-_{PMF} \approx$ 120T (Fig. 3a top). As a result, the wave-functions contain diamagnetic current loops with opposite chirality circulating on opposite sub-lattices, for opposite K valleys (Fig. 3a bottom) within each moiré cell.[28] This endows them with an orbital magnetic moment which couples to an external magnetic field. Thus, the pLLs can be labeled by three independent indices: the Chern-number, $C = \pm 1$, arising from their orbital magnetic moment, the sublattice index $A$ or $B$, and spin $\sigma = \uparrow, \downarrow$ (Fig. 3b). Sublattice-valley locking characteristic of zeroth order LLs guarantees that only four states are possible in each valley, so that if states $|+1, A, \sigma\rangle$ and $|-1, B, \sigma\rangle$ are in the K valley, then states $|-1, A, \sigma\rangle$ and $|+1, B, \sigma\rangle$ must be in the K' valley. Applying an external magnetic field produces an orbital magnetic Zeeman effect which splits the sets of $C = +1$ and $C = -1$ Chern sub-bands, creating a single particle gap at the CNP. When the Fermi-level is brought into one of these sub-bands, the exchange part of the Coulomb interaction further lifts their degeneracy resulting in the emergence of correlated insulating states at integer sub-band fillings, corresponding to integer moiré-cell fillings (Fig. 3c). The Chern-number of the filled sub-bands equals the sum of the individual Chern-numbers, and thus depends on how they are filled. As we show below, all our results can be understood within this model.

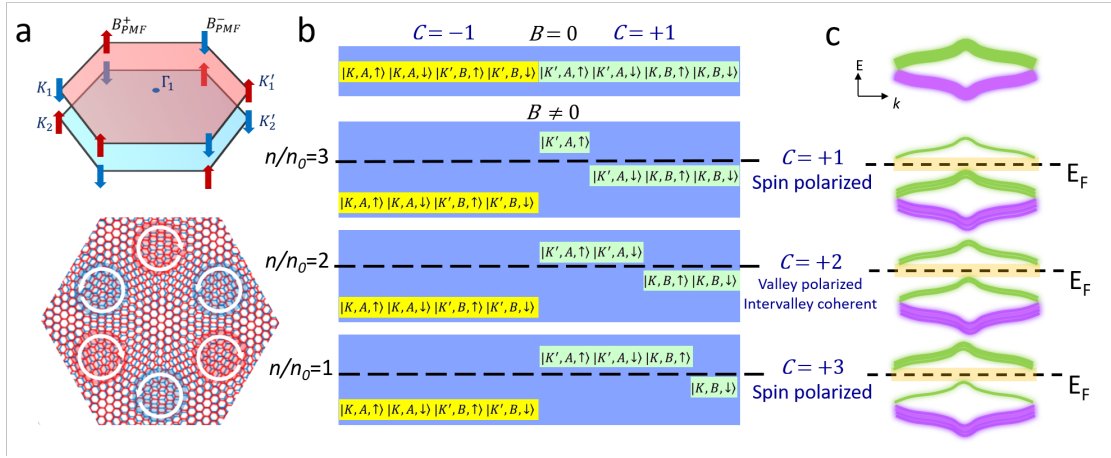

**FIG. 3 Pseudo magnetic fields, pseudo Landau Levels and Chern insulators**
**(a)** Top panel: schematic illustration of the opposite sign pseudo magnetic fields (red and blue arrows) at the K and K' Brillouin zone corners of MA-TBG. Top (pink) and bottom (cyan) hexagons represent the Brillouin zones of the top and bottom layers of MA-TBG. Bottom panel: real space current density distribution of the K valley wave-function within a moiré-cell forming circularly polarized orbital current loops with opposite chirality.[28] **(b)** The pseudo Landau levels, shown in the electron sector, are separated into two groups according to their chirality, $C = -1$ (yellow) and $C = +1$ (green), each labeled by the sublattice index A/B, valley K/K' and spin ↑/↓. For B=0 (top line) the levels are degenerate. In a finite field (bottom lines), a correlation-induced gap splits the pLL manifold separating the lower energy filled bands from the higher energy empty bands. The Chern numbers for integer fillings are indicated. In the hole sector (not shown) the filling sequence is reversed, with the $C = -1$ sub-bands being filled first. **(c)** Schematic diagram depicts the emergence of correlation-induced gaps and topological sub-bands at integer moiré fillings.

The Chern-number can be obtained from the quantized Hall-conductance $\sigma_{xy} = -C(e^2/h)$ that develops in a magnetic field. The evolution of the Hall-resistance $R_{xy}$(n,B) and of $d^2R_{xx}/dn^2$ with density and field (Figs 4a,b and ED-Fig. 7) shows clearly defined trajectories on the $s = -1, |2|, 3$, branches, which at high fields become well-quantized Hall plateaus, $R_{xy} = -(h/e^2)/C$ (Fig. 4c,d) with Chern-numbers $C = -3, |2|, 1$ respectively.[29] A similar behavior is observed on the $s = -3$ branch (ED-Fig. 8).

We first consider the states on the $|s| = 3$ branches corresponding to filling a single electron or hole per moiré-cell. In high fields, the appearance of a doping induced correlation gap, splits off a low energy (high-energy) sub-band for $s = -3$ (+3) (Fig. 3b,c). Since only one out of the eight sub-bands is occupied (empty), the state must be ferromagnetic with Chern-number $C = \pm 1$, consistent with the observed quantized Hall resistance $R_{xy} = \mp h/e^2$ (Fig. 4c and ED-Fig. 8a). The evolution of the correlation gaps with in-plane magnetic field, discussed below, further supports this scenario. It is worth comparing with reports of a $C = 1$ Chern-insulator observed at $n/n_0 = 3$, where the alignment with the hBN substrate, which breaks the $C_2$ symmetry by imposing a staggered sublattice potential,[15,17,18] is responsible for opening the gap at very low magnetic fields (~150mT). By contrast, here there is no alignment with the hBN substrate, the $C_2$ symmetry is not explicitly broken, and the $n/n_0 = 3$ state is gapless at low fields. However once the gap opens, regardless of whether it is due to the staggered potential or the magnetic field, the sub-band topology becomes apparent as evidenced by the quantized Hall resistance.

For the $s = \pm 2$ branches, two out of the eight pLLs are occupied (empty). In the presence of the magnetic field, the two $C = -1$ ($C = +1$) Chern bands, which are lowest (highest) in energy due to the orbital Zeeman effect, will be occupied (emptied) first. Therefore, regardless of which of these two bands are occupied (empty), the resulting state will have $C = -2$ ($C = +2$) as observed experimentally (Figs. 4c,d). The final state could be either a valley polarized state or an inter-valley coherent state, depending on whether the two occupied bands are in the same valley (e.g. $|+1, A, \uparrow\rangle_K |+1, A, \downarrow\rangle_K$) or in opposite valleys (e.g. $|+1, A, \uparrow\rangle_K|+1, B, \downarrow\rangle_{K'}$ or $|+1, A, \uparrow\rangle_K|+1, B, \uparrow\rangle_{K'}$) (Fig. 3b). The magneto-transport measurements are unable to distinguish between these choices.

For $s = -1$, the three electrons in this state must occupy three of the four available lowest energy $C = -1$ Chern bands. This necessarily produces a ferromagnetic state with $C = -3$, consistent with the observed quantized Hall resistance (Fig. 4b,d).

On the $s = 0$ branch where interactions with the moiré potential are reduced due to the low carrier density, the system exhibits the standard sequence of precisely quantized Hall-plateaus (Fig. 1d and ED-Fig. 3a) $\sigma_{xy} = \pm 4\frac{e^2}{h}, \pm 8\frac{e^2}{h}$, already at the lowest fields instead of the single plateau, $\sigma_{xy} = \pm 4\frac{e^2}{h}$, expected for the Chern-insulator in the pseudo LL model. This is in contrast to the $s \neq 0$ branches where only the quantum-Hall plateau of the corresponding Chern-insulator is observed. Hence we use different notations, C and ν, to distinguish between quantum Hall plateaus arising from the Chern-insulator and the LL scenario respectively.

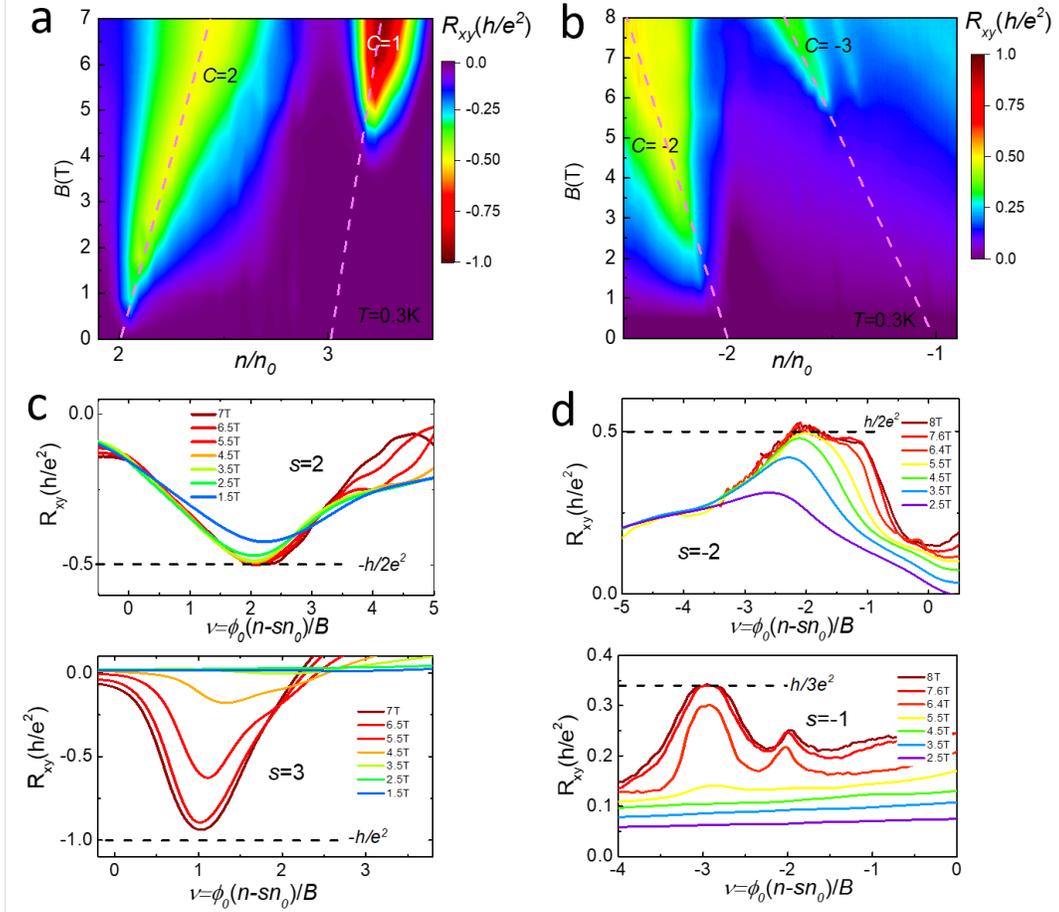

**FIG. 4 Chern insulators at integer fillings (a-b)** Evolution of $R_{xy}(n/n_0)$ with magnetic field in the conduction (a) and valence (b) bands shows the emergence of well-quantized Hall plateaus with Chern-numbers $C = -2, -1, +2, +3$ on branches $s = 2, 3, -2, -1$ respectively. **(c-d),** Field dependence of Hall resistance as a function of the flux filling factor, $\nu = \phi_0(n - sn_0)/B$, saturates to well-quantized Hall plateau values, $R_{xy} = \frac{1}{C}\frac{h}{e^2}$ corresponding to Chern-insulators with $C = -2, -1, +2, +3$ on branches $s = 2, 3, -2, -1$, respectively.

**Correlation gaps obtained from thermally activated transport**

Having inferred the existence of the correlation gaps from the doping dependence of $n_H$, we calculate their magnitude from the temperature dependence of the thermally activated part of the longitudinal resistance, $R^*$. $R^*$ was obtained from $R_{xx}(T)$ after subtracting a linear in $T$ background (SI and ED-Fig. 9).[30] Performing an Arrhenius analysis, $R^* \sim \exp(-\Delta_s/2k_BT)$, we obtained the thermally activated gaps, $\Delta_s$, where $s$ is the branch index and $k_B$ is Boltzmann's constant. To gain insight into the nature of the Chern-insulators we measure the dependence of $R^*$ on the in-plane ($B_\parallel$) and out of-plane ($B_\perp$) magnetic-fields. As shown in Fig. 5a, $\Delta_{+2}$ and $\Delta_{-2}$ decrease linearly with both $B_\parallel$ and $B_\perp$. This indicates that the insulating states on the $|s| = 2$ branches are not spin polarized (Fig. 5a bottom panel), consistent with the measured $C = |2|$ values, and can be described by either a valley polarized state or an inter-valley coherent state. The effective gyromagnetic ratios obtained from the slopes, $g_\perp = 2.1 \pm 0.3$, $g_\parallel = 2.4 \pm 0.1$ for $\Delta_{\mp 2}$, and $g_\parallel = 2.5 \pm 0.3$ for $\Delta_{-2}$, are close to the bare value, $g_0 = 2$, suggesting that the field-dependence is controlled by spin response.

Turning to the in-pane field-dependence of $\Delta_{+3}$ in Fig. 5b, we find that it increases linearly with in-plane field, with a slope corresponding to $g_\parallel = 1.5 \pm 0.2$ and a finite-field intercept, $B_0 = 1.1 \pm 0.1$T. These results suggests that the correlated state observed on the $s = 3$ branch for $B_\parallel > 2.5$T is a spin polarized insulator (Fig. 5b bottom panel), consistent with the measured $C = 1$ Chern-number, and with the theoretically predicted stripe ferromagnetic insulator phase.[31]

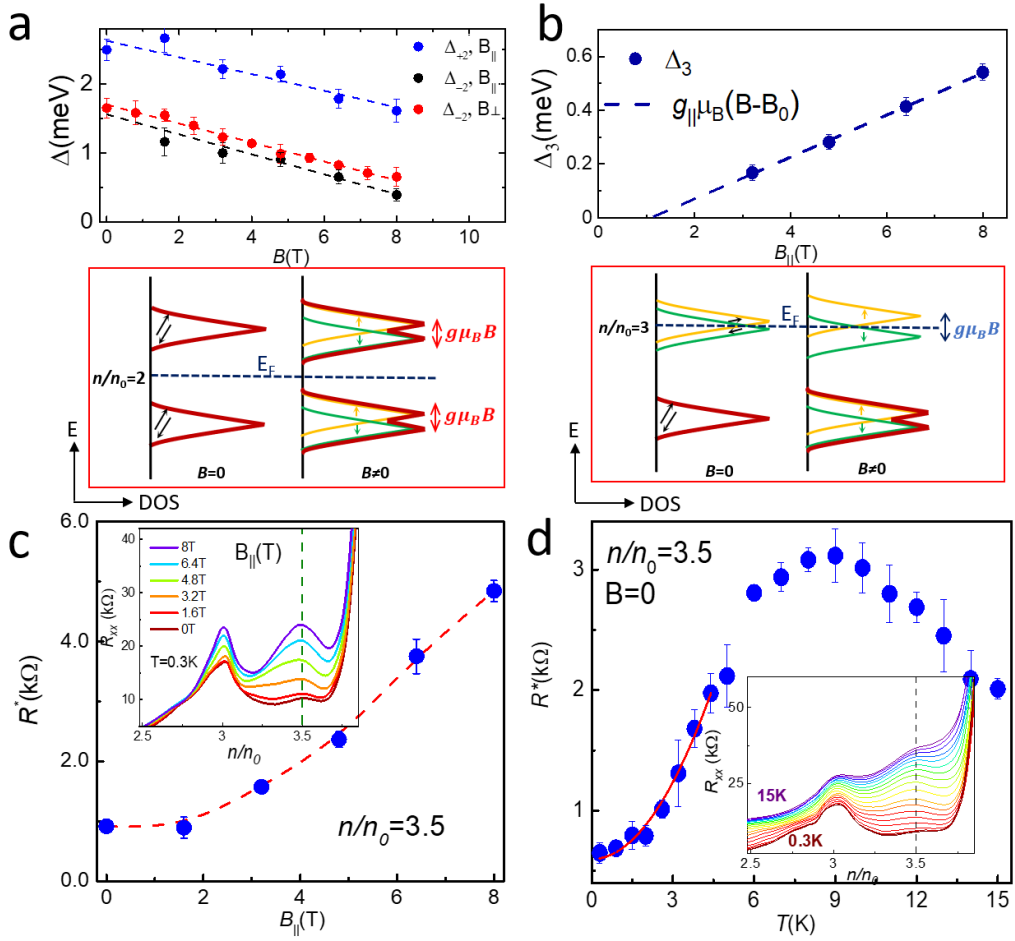

**FIG. 5 Field dependence of thermal activation gaps on the $s = 2, 3$ branches.** **(a)** Top: in-plane and out-of-plane field dependence of the thermally activated gaps, $\Delta_{+2}$ and $\Delta_{-2}$, on the $s = +2$ and $s = -2$ branches (symbols). Fits to a linear field dependence (dashed lines) indicate unpolarized insulating states. Bottom: Diagram illustrating the evolution of the gap with magnetic field. With increasing field, the Zeeman splitting in the conduction and valence bands causes the gap to decrease. **(b)** Top: in-plane field dependence of the gap on the $s = 3$ branch (symbols) and linear fit (dashed line) indicates the emergence of a field-induced spin polarized insulating state. Bottom: Schematics of field-induced Zeeman gap in the conduction band. **(c)** Field dependence of $R^*$, at $n/n_0 = 3.5$, shows a steady increase with in-plane field for $B_{||} > 2T$. Dashed line is a guide to the eye. Inset: In-plane field dependence of $R_{xx}(n/n_0)$ curves at $T = 0.3K$. **(d)**

Temperature dependence of the thermally activated portion of the longitudinal resistance $R^*$, at $n/n_0 = 3.5$ and $B = 0$, shows a crossover from low-temperature metallic, to high-temperature insulating behavior at $T \sim 9K$ suggesting an entropically driven transition. At low temperatures $R^*$ fits a quadratic, Fermi liquid-like, temperature dependence (solid line). Inset: Temperature dependence of $R_{xy}(n/n_0)$ curves in zero field.

**Surprises at 3.5 moiré filling**

In Fig. 5c,d and ED-Fig. 10 we show the temperature and in-plane field dependence of $R^*$ and the $R_{xx}$ peak which accompanies the VHS at $n/n_0 \sim 3.5$. Remarkably, as shown in Fig. 5c, $R^*$ at $B = 0$ increases with temperature, and then decreases after peaking at $T \sim 9K$. This suggests the emergence of an unusual insulating state at higher temperatures akin to the Pomeranchuk effect in $^3$He, where a fluctuating internal degree of freedom, such as nematic fluctuations, produces an entropically ordered state.[32-34] The evolution of this peak with magnetic field (Fig. 5c, ED-Fig. 10a) suggests the emergence of a gap at higher fields resulting in the formation of a new sub-band in the fractionally filled moiré-cell. This could either reflect a field-induced broken symmetry that halves the area of the moiré-cell at this filling, or alternatively a fractional Chern-insulator, but resolving this question is left to future work.

The results reported here demonstrate the creation of correlation-induced gaps at integer moiré-fillings, leading to Chern-insulators with broken flavor-symmetry and non-trivial topology. On the $|s| = 2, 3$, branches, the enhanced *e-e* interactions, enabled by the appearance of a VHS, facilitate Stoner instabilities and the emergence of Chern-insulators at relatively low fields. In the absence of such VHS on the $s = -1$ branch, higher fields are required to observe the Chern-insulator. Interestingly, the appearance of a VHS at $|n/n_0| = 3.5$ for $B > 3T$, suggests the emergence at higher fields of a correlation-induced gap and the creation of a Chern-insulator at a fractional moiré filling. This VHS is accompanied by a crossover from low temperature metallic, to high temperature insulating behavior, characteristic of entropically driven Pomeranchuk-like transitions.

**Data Availability**

The data that support the findings of this work are available from the corresponding author upon reasonable request.

**Author Contributions**

S.W., Z.Z. and E.Y.A. conceived and designed the experiment, carried out low-temperature transport measurement, and analyzed the data. S.W., and Z.Z. fabricated the twisted bilayer graphene devices. K.W. and T.T. synthesized the hBN crystals. S.W., Z.Z., and E.Y.A. wrote the manuscript.

**Acknowledgments**

We thank M. Xie, A. H. MacDonald, M. Gershenson, S. Kivelson, A. Chbukov, G. Goldstein, G. Kotliar, T. Senthil and A. Bernevig for useful discussions and Jianpeng Liu for insightful comments and discussions. Support from DOE-FG02-99ER45742 and from Gordon and Betty Moore Foundation GBMF9453 is gratefully acknowledged.

# Extended Data

## Methods

### Sample preparations and transport measurements

The hBN/TBLG/hBN/gold stacks are prepared with the dry transfer method[35] in a glove-box (Argon atomosphere), using a stamp consisting of polypropylene carbonate (PPC) film and polydimethylsiloxane (PDMS). Monolayer graphene, hBN flakes (30-50nm thick) were firstl exfoliated onto an Si substrate capped with 285nm of thermal oxidized chlorinated $SiO_2$[36]. Half of the monolayer graphene is picked up by the hBN on the stamp. The substrate with the remaining part of the graphene flake is rotated by $1°\sim1.2°$ and picked up with the hBN/G stack. The hBN/TBLG stack is then deposited onto the bottom hBN flake that is prepared separately in advance. During the assembly of the stack the temperature is kept below 160℃. The bottom hBN flake is transferred in advance onto a gold electrode as local gate and is annealed at 250°C in $Ar/H_2$ for 6 hours for surface cleaning. Atomic force microscopy (AFM) and electrostatic force microscopy (EFM)[37] are subsequently used to identify a clean bubble-free region of the TBLG prior to depositing the electrical edge contacts (Cr/Au) for transport measurements.[38]

Four-terminal resistance measurements are carried out in a He-3 refrigerator with a base temperature of 0.3K. The measurements are acquired with ac current excitation of 5-100nA, using standard lock-in technique at 13.7Hz as well as the delta-mode of a KE6221 ac current source.

### Determining the twist angle

The relation between total carrier density, $n$, and gate voltage, $V_g$, $n = 0.31 \times 10^{12} V_g (\text{cm}^{-2})$, was determined from the Landau fan and the quantized Hall plateaus at CNP. Based on this relation, we estimated the carrier density at the band edges (namely full filling $n = 4n_0$) and the value of $n_0 = 0.8 \times 10^{12} \text{cm}^{-2}$ from which we obtained the twist angle in radians $\theta = a \left(\frac{2}{\sqrt{3}n_0}\right)^{\frac{1}{2}}$, where a =0.246nm is graphene's lattice constant.

**Moiré pattern, Brillouin zone and flat band in MA-TBG** are schematically depicted in ED-Fig. 1.

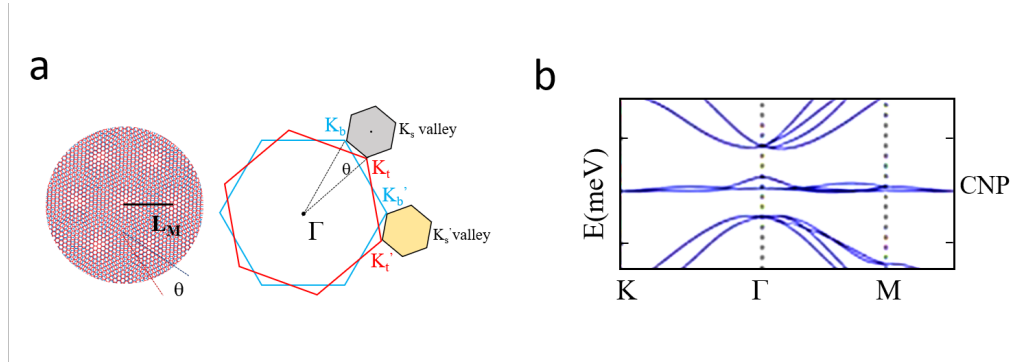

**Extended Data Figure 1 | Moiré pattern, Brillouin zone, and band structure of MA-TBG a,** Left panel: a moiré pattern with periodicity $L_M$ forms by introducing a twist angle $\theta$ between the crystallographic axes for two superposed graphene layers. Right panel**:** The moiré mini-Brillouin zone consists of two hexagons (gray and yellow) constructed from $K_s = \left|\overrightarrow{K_t} - \overrightarrow{K_b}\right|$ and $K'_s = \left|\overrightarrow{K'_t} - \overrightarrow{K'_b}\right|$ where $\overrightarrow{K_{t,b}}$, $\overrightarrow{K'_{t,b}}$ are the wave-vectors corresponding to the Brillouin zones corners of the top (t) and bottom (b) graphene layers. **b,** Schematic low-energy band structure of MA-TBG shows the flat bands near the CNP and the gaps separating them from the remote bands.

### Signatures of superconductivity near $n/n_0 = -2$

Signatures of superconductivity (SC) emerge in this device below 5K. The R(T) measurement mapping the SC dome versus doping is shown in ED-Fig.2a. Here we focus on the moiré-filling range $n/n_0 = -2.4$. In ED-Fig. 2b, a sharp resistance drop from $10k\Omega$ to $200\Omega$ is observed within the temperature range 8K-0.8K. The resistance then remains nearly constant from 0.8K to 0.3K. The temperature at which the resistance drops to half its value in the normal state, defined here as the critical temperature, is $T_c \sim 3.5K$. In ED-Fig. 2c the nonlinear current-voltage (*I-V*) characteristics in zero field, indicates a critical current of $\sim$ 12nA. In the presence of a 0.2T magnetic field at $T = 0.3K$ Superconductivity is suppressed as shown by the linear IV. The finite resistance value ($\sim 200\Omega$) at base temperature (0.3K) and the voltage fluctuations were cause by a non-ideal voltage leads which often affect four-terminal measurements of microscopic superconducting samples as reported previously.[39,40] From ED-Fig. 2d the critical field is estimated at $B_c \sim 0.02$T.

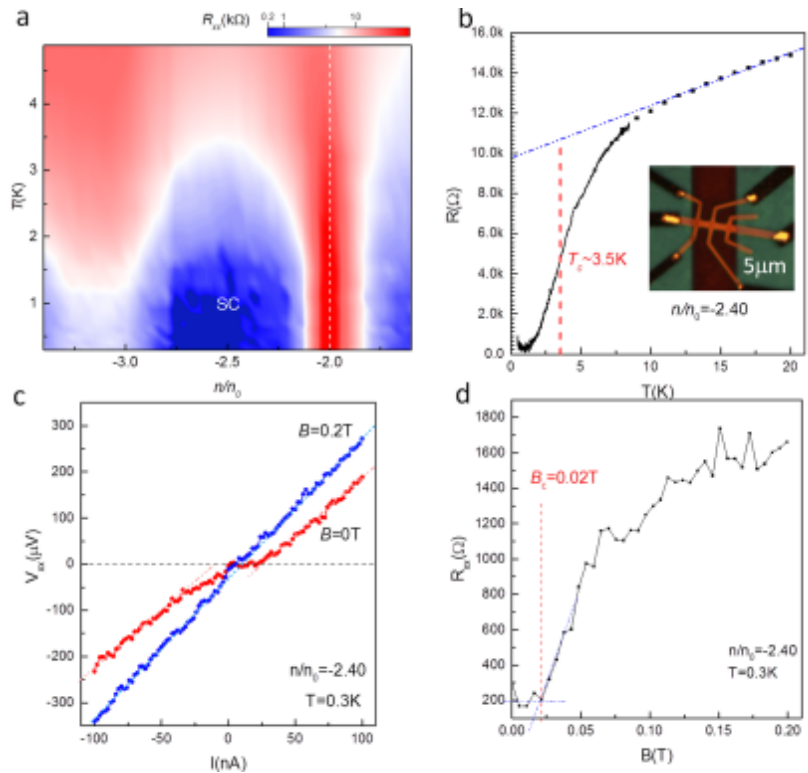

**Extended Data Figure 2 | Signature of superconductivity near $n/n_0 = -2$ a,** *R(T)* measurement mapping the SC dome versus doping. **b,** R(T) at $n/n_0 = -2.4$. The critical temperature, $T_c = 3.5$K is marked. The driving current is 10nA. Inset, Optical micrograph of the device (the brown-colored Hall bar is edge contacted (yellow) with metal electrodes (black)) **c,** Comparison of I-V curves at $n/n_0 = -2.4$ at zero and finite magnetic field shows suppression of critical current from 12nA down to zero . **d,** R(B) curve measured with a 10nA current, indicates a critical field of $B_c \sim 0.02$T.

**Evolution of quantum Hall plateaus in a magnetic field**

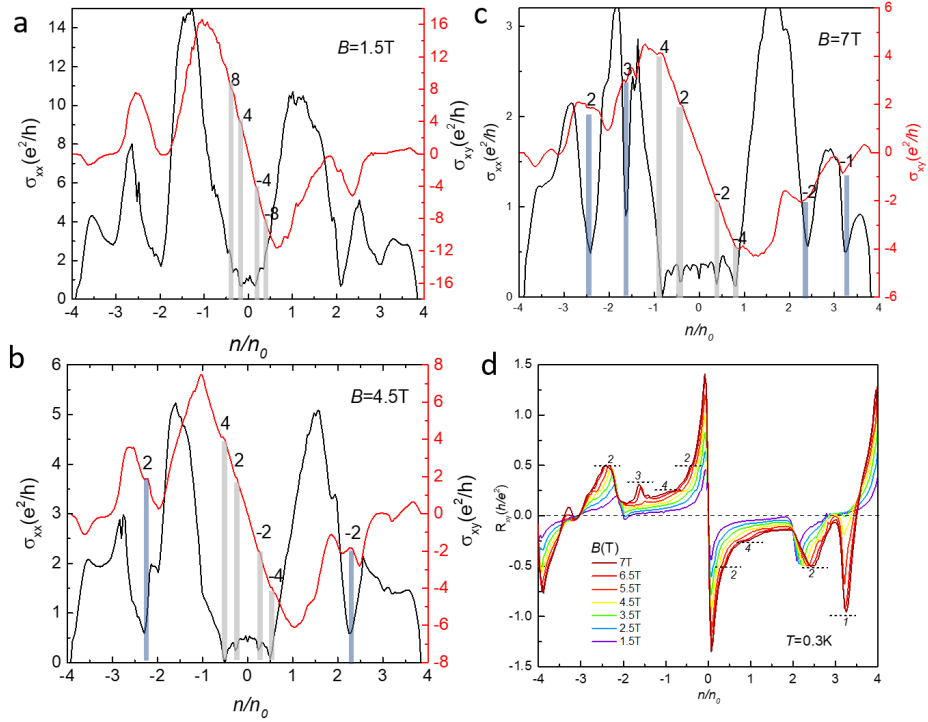

**Extended Data Figure 3 | Quantum Hall plateaus.** **a,** $\sigma_{xy}(n/n_0)$ at 1.5T displays quantum Hall plateaus, $\sigma_{xy} = \nu \frac{e^2}{h}$, and concomitant minima in $\sigma_{xx}(n/n_0)$ (gray bars) near the CNP. The Landau sequence $\nu = \pm 4, \pm 8$ indicates the 4-fold degeneracy. **b-c,** Same as panel (*a*) at 4.5T and 7T shows a new sequence with $\nu = \pm 2, \pm 4$ and concomitant minima, indicating that either spin or valley degeneracy is lifted by the field. **d,** Filling $R_{xy}(B)$ shows the emergence of Chern-insulators in the higher order branches. All data are taken at $T = 0.3$K.

**Calculating the Hall density from the Hall resistance measurements.**

The doping dependence of the Hall density, $n_H = -B/(eR_{xy})$ is shown in ED-Fig. 4a. To better resolve its intrinsic features we use the slope $dR_{xy}/dB$ of obtained from a linear fit of the measured $R_{xy}(B)$ curves at fixed $n/n_0$, as shown in ED-Figs. 4b,c. This was used to obtain the doping dependence of $n_H = -(1/e)(dR_{xy}/dB)^{-1}$ shown in Fig. 2b of the main text.

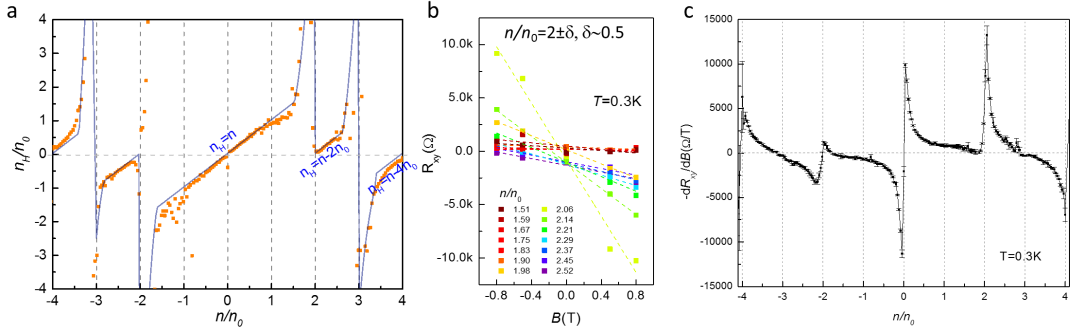

**Extended Data Figure 4 | Calculating $n_H$. a,** Doping dependence of the Hall density, $n_H = -B/(eR_{xy})$, at $B = 0.8T$ **b,** Linear fits of $R_{xy}(B)$ at fixed moiré fillings, $n/n_0$, as indicated in the legend. **c,** Filling dependence of $dR_{xy}/dB$ obtained by fitting $R_{xy}(B)$ curves as illustrated in panel (b). Linear fits of $R_{xy}(B)$ at fixed moiré fillings, $n/n_0$, as indicated in the legend.

### Estimating, $\omega_c \tau$ ,

The expression for the logarithmic divergence of $n_H$ near a VHS is valid in the low-field limit[22] $\omega_c \tau \ll 1$ where $\omega_c = \frac{eB}{m}$ is the cyclotron frequency and $\tau$ the scattering time. To ensure the validity of the fit in the main text we estimated the value of $\omega_c \tau$ as a function of density and field. Within the Drude model, $\rho_{xx} = \frac{m}{ne^2\tau}$, $\rho_{xy} = \frac{B}{ne}$, we estimate $\omega_c \tau = |\rho_{xy}/\rho_{xx}|$ shown in ED-Fig. 5. Clearly all the data taken near the putative VHSs is in the low field limit.

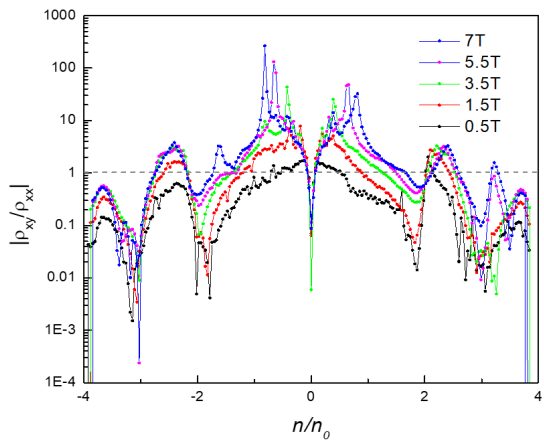

**Extended Data Figure 5 | Estimating $\omega_c \tau$** $|\rho_{xy}/\rho_{xx}| = \omega_c \tau$ as a function of $n/n_0$ at several fields as marked.

**Divergent Hall density and the VHS point near $n/n_0 = 3.5$ in high fields**

With the gap opening at $n/n_0 = 3$, we observe a divergent dependence of $n_H$ on carrier density at $n/n_0 = 3.5$. Based on the estimate of $\omega_c\tau \ll 1$ around $n/n_0 = 3.5$ (ED-Fig. 6a), the expression for the logarithmic divergence of VHS in the low-field limit, $n_H \approx \alpha + \beta|n - n_c|\ln|(n - n_c)/n_0|$ described in the main text was used in fitting the density dependence of $n_H$ in this regime, as shown in ED-Fig. 6c.

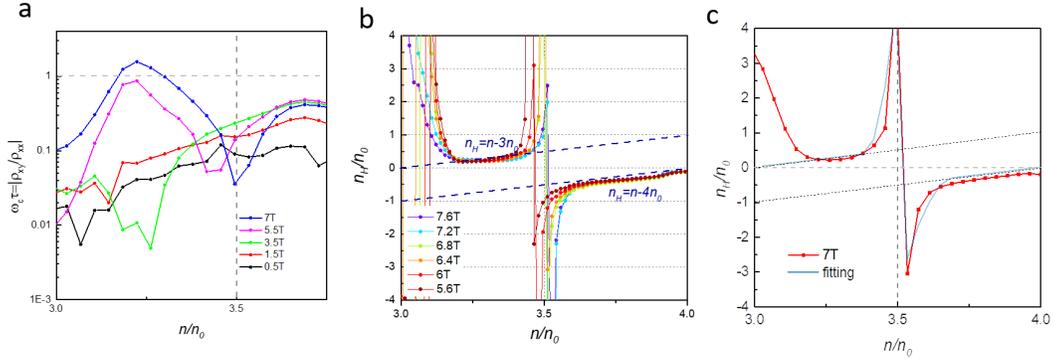

**Extended Data Figure 6| Divergent Hall density and VHS near $n/n_0 = 3.5$**
**a,** $\omega_c\tau$ around $n/n_0 = 3.5$ obtained from $|\rho_{xy}/\rho_{xx}|$ at several $B$-fields shows that the low field limit $\omega_c\tau \ll 1$ is valid in this regime. **b,** Evolution of Hall density with field near $n/n_0 = 3.5$. The divergent $n_H$ behavior is clearly resolved after the gap opens on the s = 3 branch with $n_H = n - 3n_0$. **c,** Hall density around $n/n_0 = 3.5$ fits the logarithmic divergence (solid blue line) expected for a VHS as discussed in the main text.

# Field-induced gaps and Chern insulators

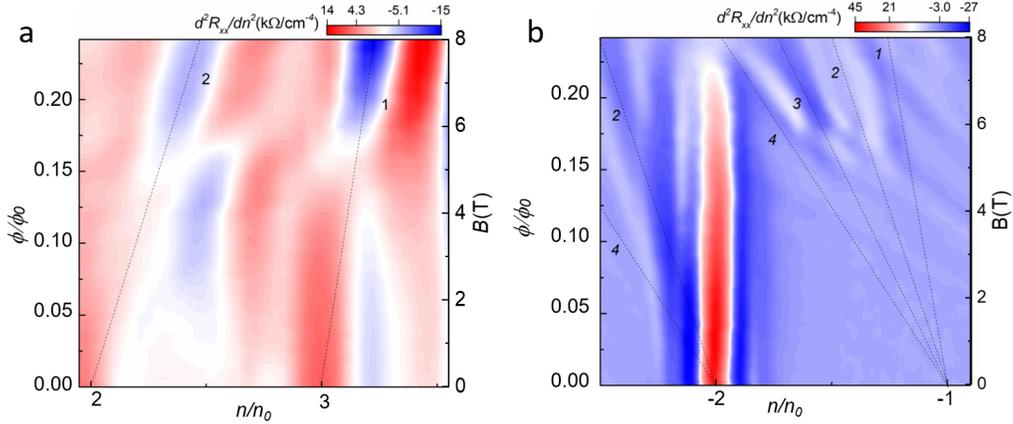

**Extended Data Figure 7 | Field-induced insulating states at integer fillings a-b,** Field and density dependence of $d^2R_{xx}/dn^2$, at $T = 0.3K$ reveals the emergence of half-Landau fans on the $s = 2, 3$ branches (a) and on the s= -1, -2 branches (b) marked by black lines.

# Quantized Hall resistance $R_{xy} = h/e^2$ in the $s = -3$ branch

In the $s = -3$ branch, quantized $R_{xy} = h/e^2$ is observed for fields above 7.2T indicating the emergence of a $C = 1$ Chern-insulator as shown in ED-Fig. 8.

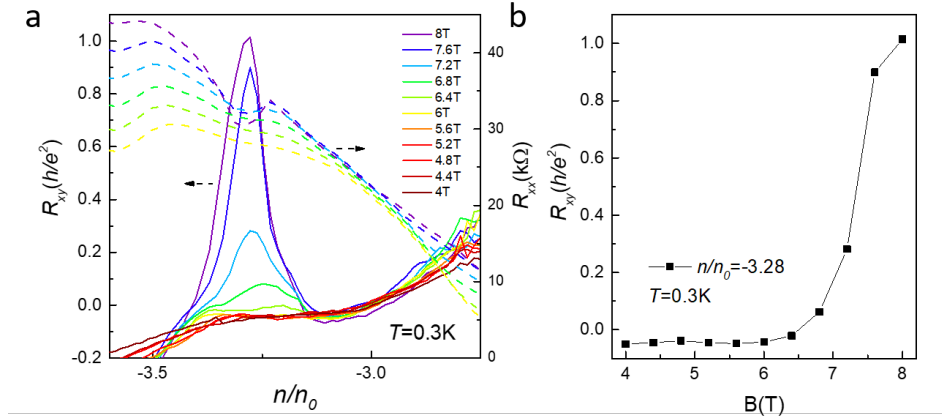

**Extended Data Figure 8| Quantized Hall resistance $R_{xy}= h/e^2$ in the $s = -3$ branch a,** Hall resistance in the $s = -3$ branch, saturates at a quantized value, $R_{xy}= h/e^2$ indicating the emergence of a $C = 1$ Chern-insulator. The corresponding $R_{xx}$ curves are shown in dashed lines. **b,** $R_{xy}$ at fixed carrier density ($n = -3.28n_0$) as a function of magnetic field shows the onset of the emergent Chern-insulator at ~6.5T.

**Subtraction of the linear-temperature-resistivity background for the estimation of thermal activation gaps $\Delta_a$ at integer fillings**

It is noted that spectroscopic gaps (e.g. at $|n/n_0| = 2$) obtained from local measurements such as STS or electronic compressibility, 7.5meV;[12] 4-8meV;[14] 3.9meV;[41] are larger than thermally activated gaps obtained from in transport, 0.31meV;[7] 1.5meV;[39] 0.37meV.[10] Such discrepancies are expected when comparing local to global probe measurements, because in the presence of gap inhomogeneity the latter are necessarily dominated by the smallest gaps.[41]

In MA-TBG, linear T-resistivity behavior is unique and prominent thus playing a crucial role in carrier resistivity. As shown in SI, the longitudinal resistance has a linear in temperature background that is observed at all temperatures and fillings. It is found that with decreasing temperature from 60K, even though the overall resistivity decreases linearly, the resistive humps at integer fillings start showing up which is consistent with previous reports.[7,10] This indicates the coexistence of the T-linear behavior and onset of the correlation gap. At high temperature, phonon-scattered (thermally excited) carriers would short out the correlation gap. In order to access the thermally activated part of the resistivity, $R^*$, we subtract the linear in T background. The thermal activation gap, $\Delta$, is estimated by fitting to the Arrhenius dependence, $R^* \sim \exp(-\Delta/2k_B T)$, in the temperature range 7K-15K, where $k_B$ is Boltzmann's constant. ED-Figs. 9a, 9b show the results at $n/n_0 = 2$ with and without background subtraction where the value of the calculated gap is $\Delta_2 = 2.5$meV and 0.1meV, respectively. The value obtained after background subtraction, 2.5meV, matches the temperature range where the resistance peak starts showing up, and is comparable to the energy scale of the spectroscopic gaps. In addition, the opening of the correlation gap at $|n/n_0| = 3$ with increasing magnetic field is also supported by the Hall density measurements. The linear Zeeman splitting relation is consistent with the spin response of the correlated states at integer fillings.

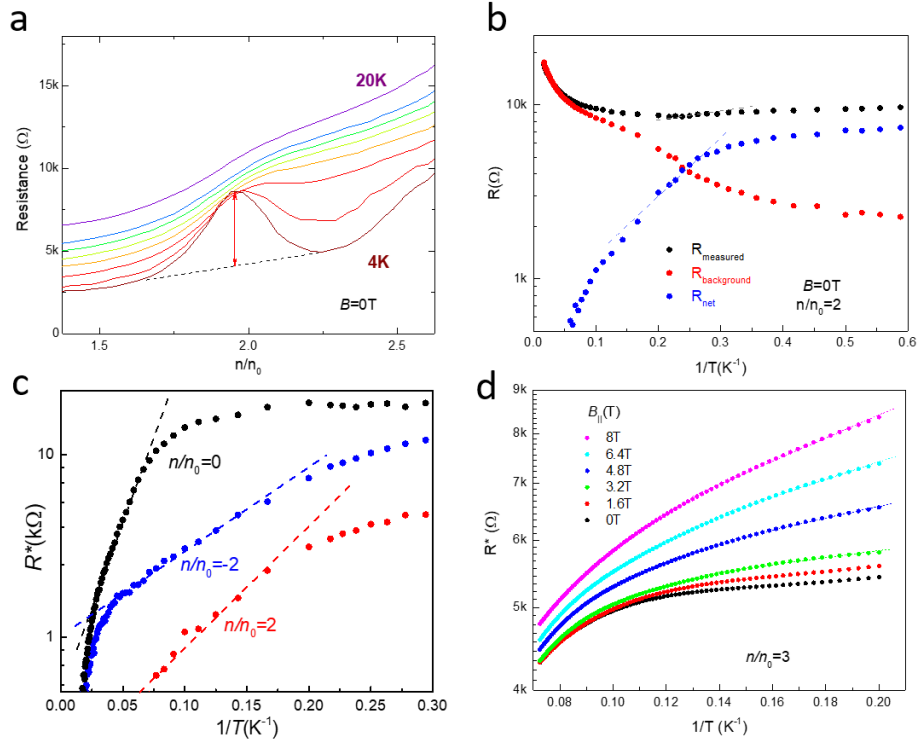

**Extended Data Figure 9| Thermal activation gaps at integer fillings and in finite magnetic fields a,** Evolution of the resistance peak around $n/n_0 = 2$ with temperature measured at $B = 0T$. The subtracted background is marked as dashed line. The net resistance ($R^*$) is marked by a red line with arrows. **b,** Temperature dependence of resistance at $n/n_0 = 2$ with and without subtraction of the background. **c**, Arrhenius fits of the temperature dependence of $R^*$ are used to calculate the thermal activation gaps at integer fillings at $B = 0T$: $\Delta_0$=7.38±0.08meV, $\Delta_2$=2.5±0.15meV, and $\Delta_{-2}$=1.7±0.15meV, for $n/n_0 = 0, 2, -2$ moiré fillings, respectively. The deviation of $R^*$ at low temperatures from the exponential divergence expected for activated transport is attributed to variable range hopping[42], which is most pronounced at $n/n_0 = 0$ where the carrier density is lowest. **d,** Evolution of temperature dependence of the net resistance, $R^*$, with in-plane field amplitude, $B_\parallel$, at $n/n_0 = 3$.

**Evolution with in-plane field amplitude of $R^*(T)$ at $n/n_0 = 3.5$;**

The miniband created by the gap opening for $B > 4T$ at moiré filling $n/n_0 = 3$ leads to a divergent Hall density at moiré filling $n/n_0 = 3.5$ that reflects the emergence of a VHS, as discussed in the main text. Another signature of this VHS is the appearance of a peak in the net resistance $R^*$. The initially positive slope of $R^*(T)$ at low temperatures, indicating metallic behavior, steadily decreases with increasing field, and becomes slightly insulating at 8T (ED-Fig. 10).

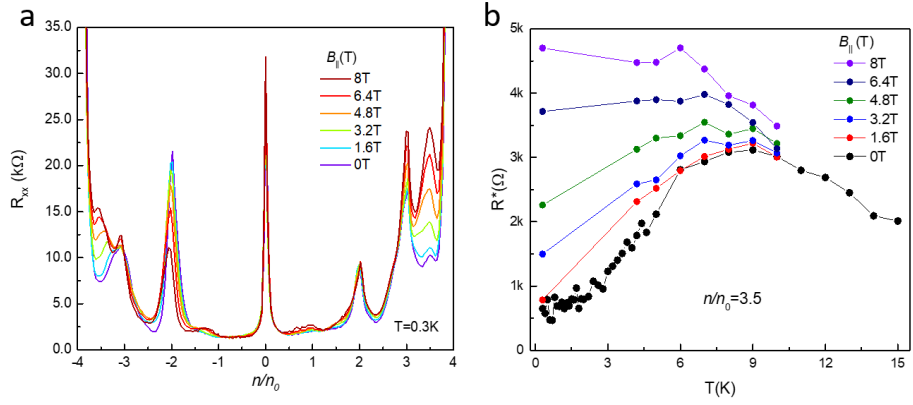

**Extended Data Figure 10| a,** Doping dependence of resistance at several in-plane fields at T = 0.3K. **b,** Evolution of $R^*(T)$ at $n/n_0 = 3.5$ with in-plane field amplitude, $B_\parallel$.

## Supplementary Information (SI)

### Shubnikov-de Haas oscillations and Berry phase.

The Landau sequence at very low field ($B = 0.4T$) exhibits 4-fold degeneracy, $\nu = -4, -8, -12$ ...as shown in Fig. S1a. The Berry phase on the $s = 0$ and $s = 2$ Landau fan branches was determined from Shubnikov-de Haas oscillations by plotting in Fig. S1c the inverse magnetic field ($1/B$) value at which the resistance minima, $\Delta R(B) = R(B) - R(0)$ obtained from Fig. S1b, occur as a function of Landau level index (LL index). Multiplying the LL index intercept of the curves in Fig. S1c,d by $2\pi$, gives the Berry phase for each branch, which is zero in both cases.

This is contrary to the zero-field non-interacting band structure calculations that find Dirac cones at the CNP[1], indicating that correlation induced gaps appear already at low fields ~0.6T. These results are testimony to the important role of interactions in this system, as well as to the high quality of the sample.

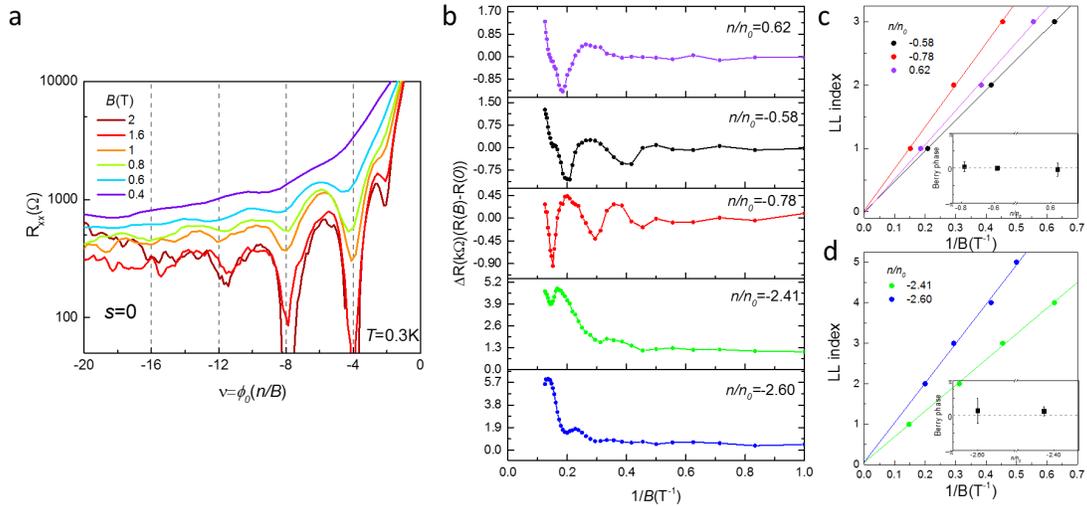

**Supplementary Figure 1| Berry phase obtained from Shubnikov-de Haas oscillations .**

**a,** $R_{xx}$ as a function of Landau filling $\nu = \phi_0(n/B)$ on the hole-side of the $s = 0$ branch at very low field ($B = 0.4 \sim 2T$). **b,** Shubnikov-de Haas oscillations near CNP and $n/n_0 = -2$. **c-d,** $1/B$-field location of Shubnikov-de Haas minima versus the Landau level index (LL index). The intercept of linear fits to the data points with the LL index axis multiplied by $2\pi$ gives the Berry phase. Insets: calculated Berry phase for both branches is zero

**Longitudinal resistivity and its dependence on temperature and moiré filling.**

The longitudinal resistivity displays linear-in temperature dependence at all moiré fillings above ~5K (Fig. S2 (a-c)). The temperature derivative in the filling range $0.5 < |n/n_0| < 3.5$, is $d\rho/dT \approx 120 \pm 20 (\Omega/K)$, which is comparable with the values reported by other groups.[2] This contribution, which is still poorly understood, has been attributed to electron-phonon interactions as well as to strange metal behavior[2].

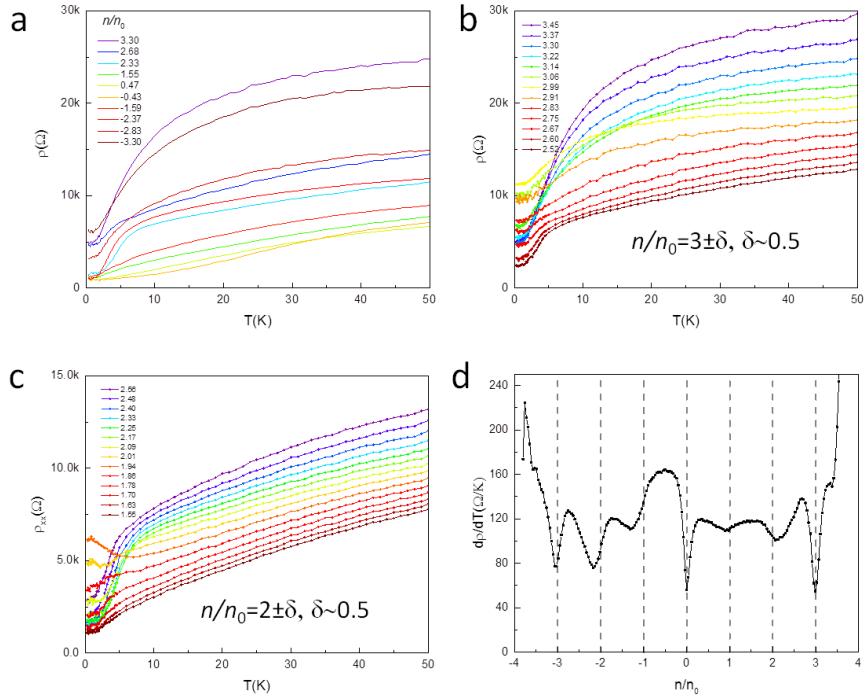

**Supplementary Figure 2| Temperature and filling dependence of longitudinal resistivity. a-c,** $\rho(T)$ curves at selected fillings. **d,** $d\rho/dT$ extracted in the linear-temperature-resistivity regions in (a-c).